# Magnetism and Superconductivity in Hydrogenated Graphite Foils

Nadina Gheorghiu

*UES Inc., Dayton, OH 45432*

Charles R. Ebbing

*University of Dayton Research Institute, Dayton, OH 45469*

Timothy J. Haugan

*The Air Force Research Laboratory (AFRL), Aerospace Systems Directorate, AFRL/RQ, Wright-Patterson AFB, OH 45433*

**Abstract**

Unique to certain unconventional superconductors is the coexistence of magnetism and superconductivity. We have previously found ferromagnetism and superconductivity in hydrogenated graphitic materials [1]. Herein we present similar as well as completely new findings this time applicable to hydrogenated graphite foils. As the strength of the magnetic field is increased, the temperature-dependent magnetization shows several important transitions: from a Neel paramagnetic-antiferromagnetic transition, to a ferromagnetic superconductor state, to an orbital paramagnetic glass high-temperature superconductor with critical temperature for the dominant phase at $T_c \sim 50 - 60$ K. The ferromagnetic state is observed up to room temperature. Thus, the magnetism of hydrogenated low-density carbon graphite foils plays an important role in establishing electronic correlations of which some are superconducting in nature.

*Keywords:* high-temperature ferromagnetic superconductor, Little model for an organic superconductor, orbital paramagnetic glass high-temperature superconductor

*Email address:* `Nadina.Gheorghiu@yahoo.com` (Nadina Gheorghiu)

*June 14, 2022*

## 1. Introduction

Until relatively recently, magnetism and superconductivity (SC) where thought as mutually exclusive states of matter. In cuprates, high-temperature superconductivity (HTSC) emerges from tuning the charge-doping of the parent phase, an antiferromagnetic (AFM) Mott insulator. The emergence of new SC materials like the (iron-based) pnictides clarified the important role played by long-range magnetic spin correlations in the coexistence of magnetism and SC on the mesoscopic scale. A ferromagnetic domain (FM) can transfer the magnetically stable spin-triplets into the neighboring nonmagnetic domains and the latter becomes SC. Furthermore, long-range proximity effects are established at the SC/FM interfaces. In addition, anisotropic materials might contain magnetically soft/hard or FM/AFM domains that are interfacially coupled, leading to the formation of quasi-Bloch walls. Under an applied reversed field, the quasi-Bloch walls are moving from within the FM domains to being compressed against the hard AFM domains. The magnetization is reversible and anisotropic [2].

Graphite, including its twisted bilayer version, can be an unconventional SCs. Starting with Dresselhaus' seminal work [3], it was soon found that FM and SC can actually coexist in graphite and other carbon (C)- based materials [4]-[6]. Notably, a lucky conjuncture of sample's properties and preparation procedure can reveal phases with $T_c$ larger than 60 K in the graphite-sulphur composites [7] and amorphous C−sulphur composites [8]. Can graphite or C−based materials reach higher $T_c$, maybe even higher than the 200 K found in a specially prepared phase of the rare-earth material YBCO [9]?

In this paper, we are bringing forth new research on magnetic and SC transitions in hydrogenated graphite foils. This is a continuation as well as a confirmation of our previous research on magnetic and SC properties of hydrogenated graphitic fibers [1, 10]. The C allotrope in this case is the graphite foil, i.e., exfoliated graphite. In pregraphitic materials such as exfoliated graphite or C fibers, the graphite layers are randomly stacked forming a turbostratic structure. This contrasts the regular stacking found in graphite crystals, where the ideal structure consists of C atoms arranged in a hexagonal/honeycomb pattern with the parallel graphene layers at $\bar{c}$ = 3.35 Å distance apart (the out-of-plane lattice constant) weakly interacting by Van der Waals forces. In-plane, each C atom is covalently bonded to three other C atoms at the distance $\bar{a}$ = 1.42 Å (the in-plane lattice constant) through $sp^2$-$sp^2$ axial hybrid orbital overlap. This layered structure of graphite leads to highly anisotropic physical properties. Ion-implantation, heat treatment, hydrogenation, or oxidation result in $sp^2$ to $sp^3$ bond conversion. Distorted $sp^2$ C bonds form grain boundaries. Importantly, disorder in the stacking sequence leads to two-dimensional (2D) weak-localization of electrons. Due to the inherent disorder, the electrons are in general confined to particular regions of the lattice, a phenomenon known as weak localization. The random potential acts as a trap for electrons, which become localized within the region of the trap. Energetically, the hopping would rather occur to an energy level close to the one for a neighboring state. The trapping results in inelastic scattering in the electron-electron (*e-e*) interactions. As 2D weak localization is a quantum phenomena, some of the *e-e* interactions can turn into SC correlations [11]. Notice also that while the σ electrons are localized, the π electrons are free to move just as conduction electrons in metals and their long-range correlations result in FM. The addition of non-magnetic (orbital) atoms like hydrogen (H) results in FM. Hydrogenated graphene was predicted for being FM [12] and this is what we have indeed found [1]. Intrinsically, graphite is already magnetic: there are AFM correlations between unlike sublattices (ABAB...) and FM correlations between like sublattices (AAA... or BBB...). In addition, disorder-induced magnetism and 2D weak-localization of electrons can lead to SC correlations as we have uncovered for oxygen-implanted C−based materials [13, 14] as well as for boron-doped C−based materials [15]. There, we have found for the dominant SC fraction that the critical temperature is in the range $T_c \sim 50 - 80$ K. Our findings agree with an earlier estimate that the shielding fraction, hence $T_c$, of metal-doped aromatic hydrocarbons increases with an increasing number of benzene rings [16]. The results presented here are in agreement with our previous work, thus confirming that hydrogenated graphite is a HTSC.

## 2. Experiment

The materials used in this work are graphite foil (Graphtek) square samples with dimensions 2 mm × 2 mm × 1 mm. Hydrogenation was done via doping by a hydrocarbon (octane, $C_8H_{18}$), resulting in what we will refer as H-C foils. The samples' mass was 7.8 mg for the C foil and 7.7 mg for the H-C foil, respectively. For comparison with results from our previous work, oxygen(O)-implanted C foils (O-C foils) were also used. The samples' O-implantation doses were $2.24 \times 10^{16}$ ions/cm$^2$ and $7.07 \times 10^{12}$ ions/cm$^2$, while their mass was 5.3 mg and 5.9 mg, respectively. The four-wire Van Der Pauw square-pattern technique was used for electrical measurements. The quality of the silver electrical contacts was checked using an Olympus BX51 microscope. Magneto-transport and magnetization measurements were carried out in the 1.9 K - 300 K temperature (*T*) range and for magnetic fields of induction *B* up to 9 T using the Physical Properties Measurement System (PPMS) model 6500 made by Quantum Design. The PPMS' sensitivity was at least $0.5 \times 10^{-6}$ emu, while samples' magnetization was 50-100% larger. The transport data was acquired using a direct current $I = 20$ μA.

## 3. Results and Discussion

Resistivity measurements for a H-C foil sample are shown in Fig. 1a, with data plotted relative to the value at $T = 300$ K. The hydrogenation decreases the C foils resistivity by one order of magnitude, $\rho_{H-C}(T = 300\text{ K}, B = 0) \simeq 72$ $\mu\Omega\cdot$cm vs. $\rho_C(T = 300\text{ K}, B = 0) \simeq 800$ $\mu\Omega\cdot$cm for the raw C foil. When a magnetic field is applied perpendicularly to the sample's area, we find: a) asymmetry in the resistance data for $B_+$ vs. the $B_-$ values; b) higher/lower $R$ for increasing values of $|B|$; c) convergence of all $R(T, B)$ values for $T \sim 250 - 280$ K, a temperature range that was found relevant in our previous work [1, 10, 13, 14]. The asymmetry in the resistance data for $B_+$ vs. the $B$ values might be a combined effect of the thermal gradient and Lorentz force on electrons vs. holes, thus leading to charge imbalance and nonlocal resistance [17]. In addition, the electrons and the holes can be confined in parallel planes resulting in the formation of a superfluid state that exists between two critical temperatures. The latter can be explained as a consequence of the electron-hole asymmetry caused by the difference in the carrier masses and their chemical potentials [18]. In [1], we have found evidence for an electron-hole pair (i.e., excitonic) superfluidity at $T_c \simeq 50$ K, where the gap in the nonlocal differential conductance was found to be practically divergent. At the same time, the fit to the temperature-dependent gap data [1] showed that the hydrogenated graphite fiber is a multigap system with some of its phases having critical temperatures above room temperature. Thus, the asymmetry of $R(B)$ is a signature for an unconventional HTSC. It is also well known that the resistance, which in in this case depends on both temperature and magnetic field, $R(T, B)$, has two components: a classical one due to the Lorentz force and a quantum one due to the 2D weak localization effect. At given $T$, the competition between the quadratic and linear terms in the field strength $H = B/\mu$ will determine the sign and the magnitude for $R(T, B)$. Thus, the magnetoresistance cannot be either symmetric or only positive as the dependence of $R$ on $H$ is not only quadratic. The in-depth analysis done on pregraphitic carbon fibers [19] can be applied here. Moreover, it is precisely the 2D weak localization effect that opens the door for SC to occur in the hydrogenated graphite [11]. One should also consider the Onsager's theorem, which states that the time reversal symmetry is broken at $H = 0$. The magnetoresistance is symmetric in $H$ while the Hall resistance is antisymmetric in $H$. The disordered nature of the graphite system here favors nonhomogeneous planar Hall effect. The interaction between the planar Hall effect and FM leads to extraordinary Hall effect and room-temperature FM [20], where the subtle role of AFM - just as it is seen here was rightfully remarked. While the H-C system does not have perfectly antisymmetric magnetoresistance, Fig. 1b clearly shows that the magnetoresistance values are determined by the symmetric-antisymmetric balance between the relevant phenomena. Another cause for this asymmetry, also mentioned in the introduction, has to do with the formation of magnetic walls [2]. The FM domains formed as of the result of octane-intercalation which leads to hydrogenation can be considered soft domains as compared to the graphitic domains that have not been modified though hydrogenation. The magnetically soft/hard FM/AFM domains are interfacially coupled and quasi-Bloch walls nucleate between the voltage contacts, more specifically, inside the FM domains. The soft FM domains have perpendicular (on the 2D layers) magnetic anisotropy. This leads to the asymmetry in the magnetoresistance after the magnetic field reversal. Notice that in the case of the excitonic SC found in the hydrogenated graphitic fibers [1], the magnitude of the Lorentz force is the same on an electron as on a hole, albeit different direction. Thus, there is no imbalance between the number of electrons and the number of holes. Interestingly, AFM and FM are sometimes competing phases, as the magnetic field can separate the spin-singlet state from the FM spin-triplet state. In the case of hydrogenated graphitic fibers, the spin-singlet was observed during cooling, while the spin-triplet was observed during warming [1]. To better illustrate the directional effect of the magnetic field, the $R(T, B = \text{fixed})$ curves (Fig. 1 a) were replotted as $R(T = \text{fixed}, B)$ curves (Fig. 1b). For temperatures close to the Neel transition temperature, 40´ K < *T* < 50 K (see later, $T_N \simeq 44$ K), the magnetoresistance is linear in the high magnitude *B* < 0 fields. This also proves that the SC occurs at the interfaces of small FM misaligned domains. The applied magnetic field moves the Bloch walls from within the FM domains towards the interfaces between these domains, resulting in spin-triplet SC coexisting with FM at these interfaces. Thus, $T_c$ can be significantly higher than in the bulk and the topology of the system is revealed by its flat-band energy spectrum, i.e., a divergent density of states [1, 21].

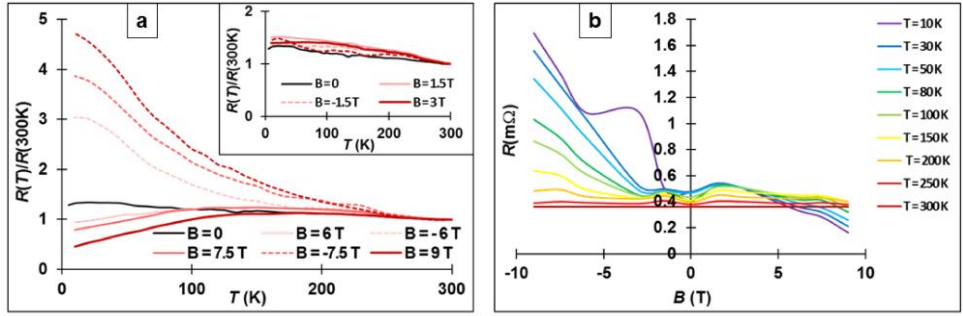

Figure 1: a) Temperature-dependent resistance (normalized to the value at 300 K) for a H-C foil in perpendicular magnetic field. The inset shows that lower *B* has a less significant effect *R*. b) The field-dependent resistance shows the asymmetry between the $R(B_+)$ vs. the $R(B_-)$ lines.

Few more remarks on the spin-triplet state: Due to the antisymmetric nature of Cooper pairs, there is either spin-singlet SC with $s-$ and $d-$wave symmetry or triplet superconductivity with $p-$ and $f-$wave symmetry. In this case, magnetic fluctuations and SC can coexist because the Cooper pair could have a non-zero magnetic moment that couples with other magnetic degrees of freedom in the system. Preliminary data that we have not published yet suggest $s-$ and $p-$wave symmetry for hydrogenated graphite. As shown, the H-C system is a spin-triplet ferromagnetic superconductor (FMSC), i.e., FM can coexist with the spin-triplet state, not spin-singlet state. After all, the FM brought into the graphite foil system by hydrogenation with octane is inhomogeneous, thus spin-triplet is likely. The likelihood of spin-triplet was more extensively discussed for the case of hydrogenated graphitic fibers [1], also a FMSC system. The pronounced zero-biased peak in the nonlocal differential conductance for the hydrogenated graphitic fiber [1] is a signature of spin-triplet SC, moreover, Andreev states were also found. Similar phenomena are expected for the hydrogenated graphite foils here. Also, while the Neel´ transition is observed for the hydrogenated graphite foils and not for the hydrogenated graphitic fibers, in both cases the magnetic field application reveals AFM, FM, paramagnetic (PM), and SC behavior, thus bicritical/tricritical/tetracritical behaviors are also expected. Moreover, the application of a high magnetic field on the H-C system can lead to the formation of a special SC state known as the Fulde−Ferrell−Larkin−Ovchinnikov (FFLO) state that has been observed in other SC layered organic materials. The relevance of the FFLO state for the hydrogenated system was at large discussed in [10].

As known, the coefficient of thermal expansion along the $c$-axis (normal to the graphitic planes) is positive and in absolute value larger than the in-plane coefficient of thermal expansion (negative), both of the order of $10^{-6}$/K. In C fibers and in graphite, the former is about 3× larger, while in graphite foil it is about 50× larger. Thus, heating of the sample results in an increase of the inter-layer distance and a shrinking of the layers. On the other hand, a transverse magnetic field results in a corresponding pressure applied on the 2D material, enhancing the anisotropy in the electronic transport. At the maximum field induction here, $B = 9$ T, the corresponding magnetic pressure is $p_{mag} \simeq 32$ GPa, which effect is unknown considering that the system is governed by Van der Waals weak interaction between its layers. What is true is that the magnetic strain leads to charge localization, thus is can lead to $e$-$e$ interactions of which some are SC with a flat-band energy spectrum. The observation of periodic strain profiles in twisted multilayers is being explained as due to the existence of a periodic pseudo-magnetic field [22]. Relevant here, the resistance upturn for fields $B_c \leq 1.5$ T is a sign of SC behavior. Interestingly, the upper critical magnetic field for $Bi_2Se_3$ whiskers doped by Pd was found to be $B_c \simeq 1.5$ T [23]. Without staggering and with a lower inter-layer distance, the graphite foil (or graphite, for that matter) would be SC just as $MgB_2$ is. The electronic similarities between graphite and other systems like $MgB_2$ or $Bi_2Se_3$ are described within the resonating-valence-bond model [24]. While not all 2D materials are SC, there is an interesting link between graphite and $Be_2Se_3$: the Pauli PM of the Cooper pair [25]. While the $Be_2Se_3$ system shows spin-triplet nematic SC, the hydrogenated graphite might be an orbital PM glass HTSC, as it will be shown below. It was found that in $Be_2Se_3$, correlations between the electron-electron, electron-phonon, and SOC interactions lead to weak antilocalization (i.e., the growth of conductivity). In the case of partial SC, there is coexistence between Kondo effect, weak antilocalization, and SC. Thus, the $R(T)$ dependence might be a combinations of weak electron antilocalization and SC.

We have also conducted magnetization measurements using the PPMS vibrating sample magnetometry option. The temperature-dependent magnetization data is shown in Fig. 2. We are looking for the Meissner effect, arguably the most important property of a SC that generally implies zero resistivity. As known, at the cooling of a SC placed in a magnetic field, the initially trapped magnetic flux is expelled when the temperature goes below the SC transition temperature $T_c$. As Fig. 2 shows, upon incrementally increasing the strength of the magnetic field $H$ from 0 to 150 Oe, several important magnetic transitions occur in the H-C foil: from reversible to irreversible PM (Fig. 2a-c), to slightly irreversible PM-AFM Neel transition (Fig. 2d-f), to reversible diamagnetism (DM) (Fig. 2g), and´ further more to irreversible DM (Fig. 2h-i). We find for the Neel transition temperature´ $T_N \simeq 44$ K (Fig. 2d-f). Significantly, when the strength of the magnetic field is high enough, the Meissner fraction becomes dominant and the H-C foil becomes SC below $T_c \simeq 57$ K (Fig. 2h-i). In addition, we observe low-$T$ upturn of the zero-field cooled (ZFC) and the field-cooled (FC) curves. This reentrant PM was observed in other high-$T_c$ systems [26] and attributed to the magnetization of the SC coming from three sources: (a) the DM shielding moment; (b) trapped flux; (c) a PM contribution (positive magnetic susceptibility $\chi_m$). The Meissner fraction depends strongly on the strength of the applied field. In the high-field limit, the trapped flux fraction can be comparable to and eventually can cancel out the DM shielding fraction. This kind of Meissner effect, PM by nature, is due to the trapping of magnetic flux [27]. On decreasing $T$, the magnetic flux captured at the third (surface) critical field inside the SC sheath compresses into a smaller volume, thus allowing extra flux to penetrate at the surface [28]. The PM Meissner effect is a general property of a small SC that is being enhanced in a plate geometry, such as our graphite foil here. From all the evidence, the HTSC occurs in the H-C materials at the interfaces, i.e., the weak SC links are formed across the grain boundaries and/or across defect surfaces within the grains. Flux trapping in the grains of a granular HTSC is the main mechanism for hysteresis, indeed, not pinning. At the S/F interfaces between a *s*-wave SC and a FM metal, the spin-orbit coupling (SOC) induces spontaneous vortices (in the absence of a magnetic field) that are pinned along the edges of the FM with the SOC reinforcing the SC state [29]. Moreover, in granular SC (which are type II SC), the occurrence of the PM Meissner effect proves the existence of π Josephson-coupled SC domains. Relevant to our system here, the PM Meissner effect, known also as the Wohlleben effect [30, 31], was found to be inherent to granular SC [32]. We also notice that in small topological (unconventional) SC, granular SC in particular, the intrinsic inhomogeneity can result in PM because of the odd-frequency Cooper pairs at the surface of the small SCs that are accompanied by zero-energy surface Andreev bound states [33]. Indeed, as we have found in [1] for hydrogenated graphitic fibers, the gap at $T_c$ = 50 K suggested interference of chiral asymmetric Andreev edge states and crossed Andreev conversion. The reentrant PM can be clearly seen to dominate over DM below a temperature $T_p \leq 20$ K. Also, when Andreev bound states appear at the interface, the direction of the induced magnetization is opposite to that without Andreev bound states [34]. A singlet-triplet spin conversion results in the sign change of the magnetization. In the Cooper pair picture, the spin structure of the dominant Cooper pair determines the direction of the induced magnetization. Notice also that in magic-angle twisted bilayer and trilayer graphene there is a superposition of singlet and spin-triplet pairing [35]. The reentrant SC is observed in the trilayer case as well as in the hydrogenated graphite here. While the bilayer and the trilayer graphene are engineered to be relatively twisted by a magic angle, the formation of the nearly flat band can be the result of a spontaneously formed, therefore natural, misfit dislocation array at graphites interfaces and the topological origin of this flat band, which can host both FM and HTSC states, can be understood in terms of the pseudo-magnetic field created by strain [36]. It is this natural, yet magical, misfit of alignment of graphite plane at the interfaces that leads to HTSC, more so than the *e-e* correlations. Just recently, it was found that several-times-fold slabs of disordered graphite obtained by deintercalation of $KC_8$ show all possible twist angles, with some of them qualifying to small magical angles such that anomalies in $R(T)$ at $T_c$ were observed even above room temperature [37]. Their results, which also found FM above 400 K, are in the same category of findings as ours [1, 10], thus confirming that disordered C-allotropes with grain sizes 50-100 nm can show both FM and a relatively small fraction Meissner SC.

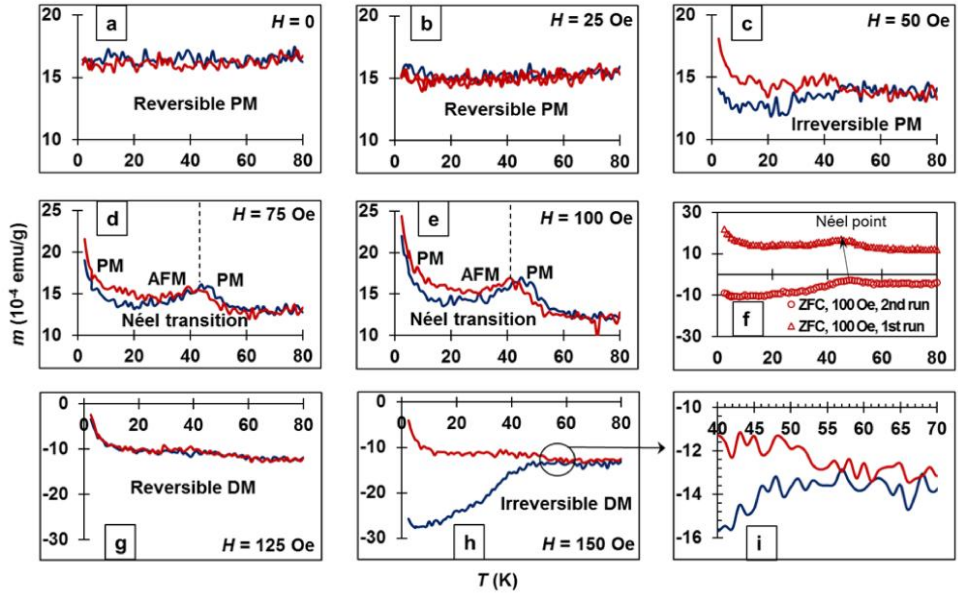

Figure 2: ZFC-FC low-field data for the H-C foil. As the strength of the magnetic field *H* is increased, several magnetic transitions are observed: reversible to irreversible PM (a-c), to slightly irreversible PM-AFM Neel transition (d-f), to reversible DM (2g), and finally to irreversible DM (h-i).

We also observe that the PM is actually metastable. At a second run under the same magnetic field strength, the magnetization is negative, and a cusp left from the Neel point can still be seen, though at a higher value, $T \simeq 49$ K (Fig. 2f). The change in magnetization sign from positive to negative shows that the trapped magnetic flux was removed from the system, which is now being driven into the SC state, apparently a more stable state than the PM one. After all, the graphite foil has irregular zig-zag edges and thus it can host PM, FM, AFM, DM, and even SC. The only way to accommodate AFM, PM, and SC is to have a FM that is also a SC, i.e., a spin-triplet FMSC.

Upon further increase in the strength of the magnetic field $H$, the magnetization becomes again positive and the PM-AFM Neel transition reemerges at the´ same transition temperature $T_N \simeq 44$ K (Fig. 3). Fig. 3b shows the Wohlleben effect mentioned before. An increase in the strength of the magnetic field has a direct effect on the intergranular currents, which contribution to the magnetization is either PM or DM. Thus, in higher magnetic fields, a reentrant PM order emerges alongside the SC order. As Fig. 4 shows, the system is not a glass, or the irreversible temperature $T_{irrev}$ would need to vary linearly with $H^{2/3}$ accordingly to the Almeida-Thouless scaling law. The behavior at magnetic field strength $H \sim$ 150 Oe (upper line above $T_c \simeq 57$ K) is similar to the $H - T_c$ line in [26] for a granular SC. The reentrance of PM and of the PM-AFM Neel transition can´ be explained by considering that the H-C foil is an example of orbital PM glass HTSC, a true new state of matter in which one can find the coexistence of orbital PM order and SC order without the Meissner effect [38]. After all, interlayer interaction in graphite can lead to enhanced PM orbital effect. The orbital PM glass is like a crystal with chaotically distributed dislocations of different types such that even a small magnetic field can polarize the system by inducing different orbital moment-flip processes. The strength of the chaotically distributed circular currents can decrease and increase in time or even change direction (i.e., have orbital moment flips). The rate of change for these currents will depend on the local dissipation in the weak links that is due to impurities and the local $H$, which in turn will depend on the values of other orbital currents. The SC regions form thick rings of Josephson junctions and the magnetic flux going through the interior of the SC rings is a chaotic line (linear, circular, other), i.e., it is a topological object. Thus, the system is an orbital PM glass and a SC at the same time, moreover, it can be a HTSC. In addition, granular disordered HTSC might have the sign of some of their Josephson loops reversed when π-contacts are being created between SC grains by the presence of magnetic impurities. In the H-C foil here, itinerant FM is introduced in the system by octane with its freely moving $H^+$ (protons) on the graphite s interfaces. Thus, the Josephson junction SC (JJSC) loops might have π-contacts. The formation of the orbital PM glass is conditioned as follows: a) the JJSC rings with an even number of π-contacts give constant negative magnetic susceptibility, i.e. conventional Meissner response, while b) the JJSC rings with an odd number of π-contacts give (positive) PM susceptibility, which is proportional to $1/H$. Clearly, $\chi_m \to 0$ as $H \to 0$ (abnormal response). At any number of JJSC rings there will exist a small $H$ for which the rings will give the main contribution to $\chi_m$, which will be PM. The systems behavior is of a magnet with local orbital magnetic moment. On the other hand, the magnitude of the anti-Meissner signal decreases with increasing $H$ and, for sufficiently large $H$, $\chi_m < 0$, i.e., the DM Meissner effect. As Figs. 2-3 show, SC dominates over PM for 100 Oe < $H$ < 200 Oe. The rather small magnetization observed for $H \sim 175 - 200$ Oe (Fig. 3c) suggests possible clustering of superparamagnetic (SPM) order for field strengths smaller than needed for the reentrant PM that eventually will coexists with a SC order for $H \sim 300$ Oe. An intermediate SPM order can be seen as transitory form of magnetism proceeding the reentrance of orbital PM glass coexisting with SC. A persistent AFM/FM background for temperatures up to 50-60 K [1], which are coming from the free $H^+$ protons, might favor the occurrence of different magnetic orders, as well as the orbital glass behavior.

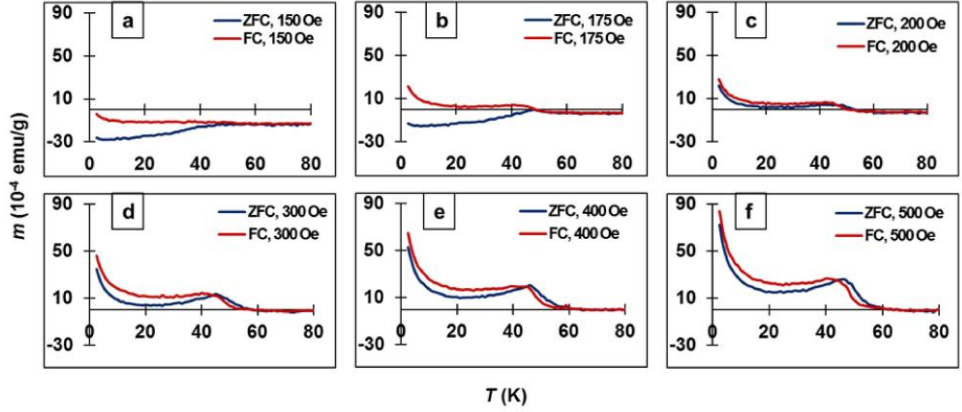

Figure 3: ZFC-FC high-field data for the H-C foil. For comparison with Fig. 2, the $H$ = 150 Oe data was again included. With the strength of the magnetic field $H$ further increased, several magnetic transitions are observed: irreversible DM (3a), to irreversible PM-DM (3b), to practically reversible PM (3c), to slightly irreversible PM-AFM Neel transition (3d-f).

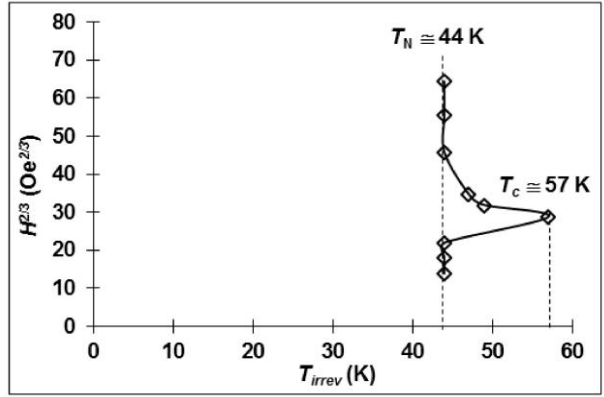

Figure 4: The H-C foil is not a glass system, as $H^{2/3}(T)$ is not an Almeida-Thouless line. The reentrance of PM followed by more PM-AFM Neel transitions suggests that the H-C foil might be´ an orbital PM glass HTSC, where orbital PM order and SC order coexist without the Meissner effect.

For moderate field strengths $H \geq 200$ Oe, the PM dominates. We observe both reentrant PM and more PM-AFM Neel transitions in the H-C foil, the latter occur-´ring at the same $T_N \simeq 44$ K (Fig. 3d-f). To our knowledge, these are completely new features, for which a new mechanism would be needed to explain both the coexistence of SC and orbital glass PM as well as the reentrant features observed in the H-C system. In high magnetic fields, the SC and the PM orbital glass are decoupled when the field direction is reversed and either DM ($\chi_m < 0$) or PM ($\chi_m > 0$) wins (Fig. 1).

For comparison, the ZFC-FC data for raw foil (C-foil), O-implanted (O-foil) at two doses (max/min), and C foil hydrogenated by octane treatment (H-C foil) are shown in Fig. 5. The irreversibility temperatures $T_c$ are: 55 K (raw C foil), 57 K (H-C foil), 45 K, and 50 K (O-C foil, for minimum and maximum O-implantation dose), respectively. Significantly, the H-C foil is more DM than the C foil, suggesting that the H-C foil is SC too. On the other hand, the O-C foils are SC below a $T_c$ that is lower than for the H-C foil and also lower for a smaller O-implantation dose. At the same time, higher O-implantation dose resulted in more PM and thus positive magnetization (compare the O-C foils). In any case, the values found for the critical temperature $T_c$ for either the H-C foils or the O-C foils are close to those that we have previously found for other H-C [1, 10] and O-C samples [13, 14], respectively. These values are all close to the mean-field $T_c$ for SC correlations in metallic-H multilayer graphene or in highly oriented pyrolytic graphite (HOPG), $T_c \sim 60$ K [39]. Interestingly, the remanent magnetization for the raw C foil (Fig. 5, bottom plot) clearly shows that the critical temperature $T_c \simeq 55$ K separates the low-$T$ PM orbital glass from the high-$T$ FM background FM. Note that SPM, which we have found before [15], can lead to large magnetocaloric effect and giant or even colossal magnetoresistance [13], which have significant technological applications to magnetic refrigeration or to the use of spin-injection for achieving dissipationless (using spin-triplet FMSC) long-distance transport and reading/writing of information in the new field of spintronics and topological quantum computing. Thus, one can infer that SPM is one of the dynamic features of C-based materials. Likely, the PM-SPM-PM transitions occur while magnetic domains change their volumes as the strength of the magnetic field is increased.

High-field magnetization loops $m(B)$ at temperatures $T = 2.5$ K and $T = 300$ K are shown in Fig. 6. In order to separate the SC/FM response from the huge DM background, the sample's DM response to a high field ($B = 1$ T, inset) was subtracted from the initial data (see inset). In high magnetic fields, the magnetization loops are lines and show practically no hysteresis, thus reflecting the internal DM of the graphite samples. Thus, the linear $m(B)$ data for high values of $B$ was subtracted from the m(B) data for low values of $B$. Several important features we observe with these $m(B)$ loops: a) they have both FM and DM trends for all temperatures below 300 K, suggesting that the Curie temperature for these H-C foils is higher than the room temperature; b) after the subtraction of the DM background, they have an oscillatory dependence on the field, confirming the existence of the PM Meissner effect in these H-C foils and surface SC manifested as a metastability that is due to the coexistence of multi-quanta vortex states ($L\Phi_0$,

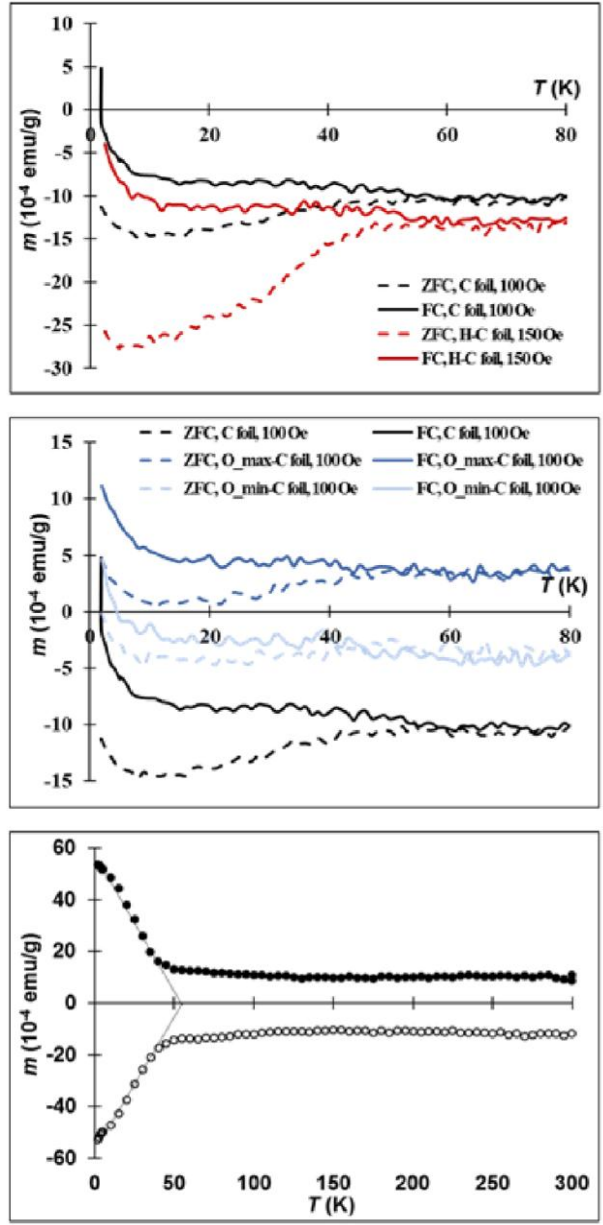

Figure 5: Upper and middle plots: ZFC-FC data for the H-C foil (red) vs. the O-C foil (light and dark blue for the two O-implantation doses (max/min), respectively). The raw C foil data is also shown (in black). Bottom plot: Remanent magnetization for a raw C foil sample obtained after the application of a field of induction $B_+ = 9$ T (filled symbols) and $B_- = -9$ T (empty symbols), respectively.

$\Phi_0 = h/2e$, $L > 1$) and single quantum ($L = 1$, or Abrikosov) vortices [40]; c) as $T$ goes up, the FM component increases at the expense of the PM component.

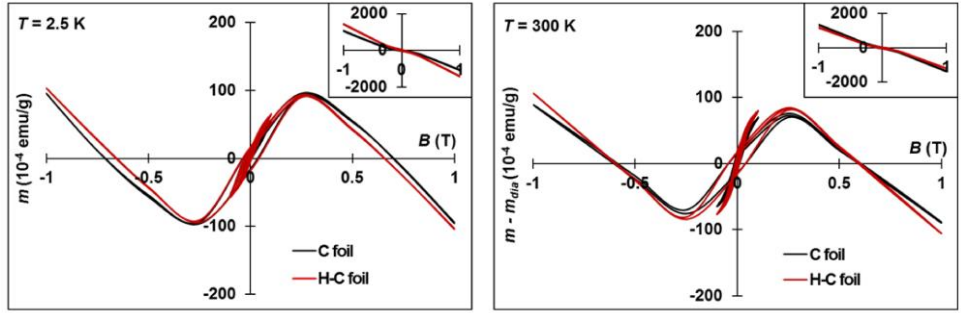

Figure 6: High-field magnetization loops $m(B)$ for $T = 2.5$ K and $T = 300$ K, respectively. Legend: raw C foil in black and H-C foil in red. DM background included (left) vs. DM background (inset) subtracted (right).

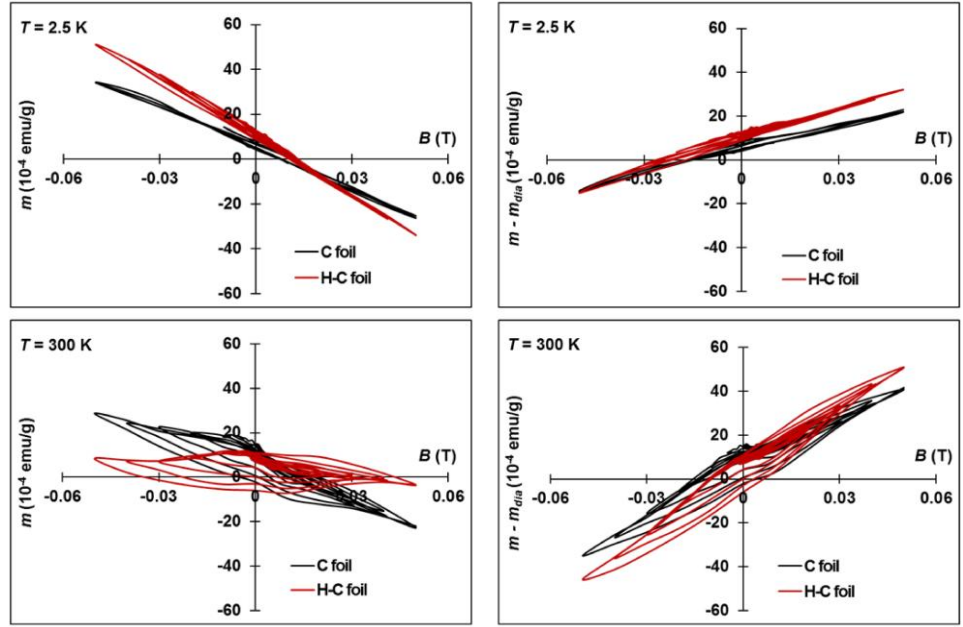

Figure 7: Low-field magnetization loops $m(B)$ for $T = 2.5$ K and $T = 300$ K, respectively. DM background included (left) vs. DM background (inset) subtracted (right). While both FM and SC observed, the hysteresis decreases at lower temperatures. The asymmetry of the magnetization loops reflects the granular nature of the material.

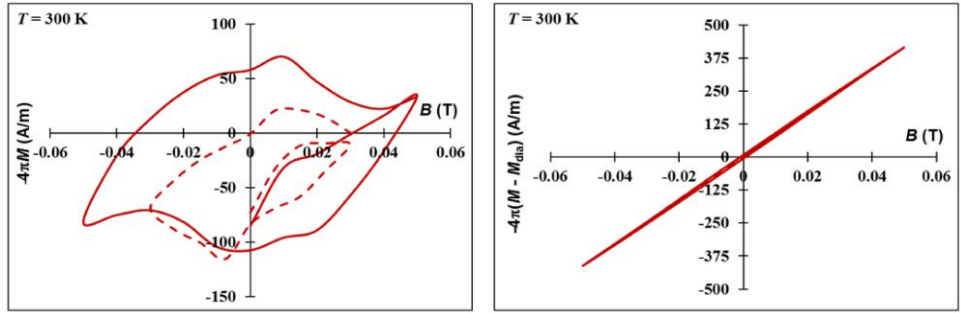

Figure 8: Magnetic moment at $T$ = 300 K with (left) and without (right) the DM background.

Low-field $m(B)$ loops are shown in Fig. 7. Here too we notice several important features: a) At low $T$, the loops are not hysteretic, a characteristic of soft materials such as graphite foil. b) Both the FM and the SC, or the FMSC, are pronounced in the H-C foil as compared to the C foil. c) The $m(B)$ the loops show the "fishtail anomaly" (kink) feature characteristic to granular SC [41]. d) The hysteresis increases with the temperature and the FMSC becomes more evident beyond the low-$T$ PM.

A couple of low-field volume magnetization loops $M(B)$ are shown in in Fig. 8. The known conversion factor was used: 1 emu = $10^{-3}$ A·m$^2$. At $T$ = 300 K, the sample is clearly FM. The absence of points on the magnetization loops that should correspond to $B_{c1}$ and $B_{c2}$ is due to the fact that the SC DM was subtracted, thus leaving out the FM behavior. Magnetic hysteresis comes mainly from the DM component (including the SC component. For the inner loop, the critical values for the magnetic induction of this type II SC is of the order of few tens of mT, $B_{c1} \simeq$ 10 mT and $B_{c2} \simeq$ 30 mT, respectively. We use these values to estimate the penetration depth $\lambda \sim$ 250 nm and the coherence length $\xi \sim$ 100 nm, respectively. The ratio $\kappa = \lambda/\xi \simeq 2.5 > 0.42$ tells us that there is a SC surface layer for which a third maximum field $B_{c3}$ can be defined [42] such that $B_{c3} = 2.4\kappa B_{c1} = 1.7 B_{c2}$. Here, we find then $B_{c3} \simeq 51 - 60$ mT. It should be pointed out that for ideal samples, the nucleation of SC regions is energetically favored to start from the samples' surface. In non-ideal samples like the ones here, the nucleation of SC regions in decreasing field is rather initiated from volume defects. It is possible that a bulk SC state (hugely-gaped) is found at $T_c \simeq 50 - 60$ K, while surface (topological) SC states (weakly-gaped), which are protected by the time-reversal symmetry, are found at all other temperatures. The hugely gaped (bulk) excitonic state at and the $T$-dependence of the SC gap found in [1] showed that the hydrogenated graphitic system can host both bulk (characteristic to Bernal stacking) and flat-band surface states (characteristic to rhombohedral stacking).

In order to separate the SC component from the total DM component, the magnetization data was replotted in two more ways: 1) The difference between the magnetic moment at $T$ = 2.5 K and the one at $T$ = 300 K for both the H-C foil and the C foil (Fig. 9). As seen, while both the H-C foil and the C foil are more DM at $T$ = 2.5 K than at high $T$ = 300 K, only the H-C foil shows a SC-like magnetization loop and also the kink feature. The loop also starts from the origin, suggesting that the offset seen with the magnetization loops is temperature-independent, thus FM. On the contrary, the offset in magnetization is not canceled for the C foil, which is PM in a field. 2) The magnetic moment of the C foil was subtracted from the magnetic moment of the H-C foil (Fig. 10). Notice that at $T$ = 2.5 K, the H-C foil is more DM than the C foil, while at $T$ = 300 K the H-C foil is more FM than the C foil. Thus, the H-C foil is a FMSC, as already shown by Fig. 7.

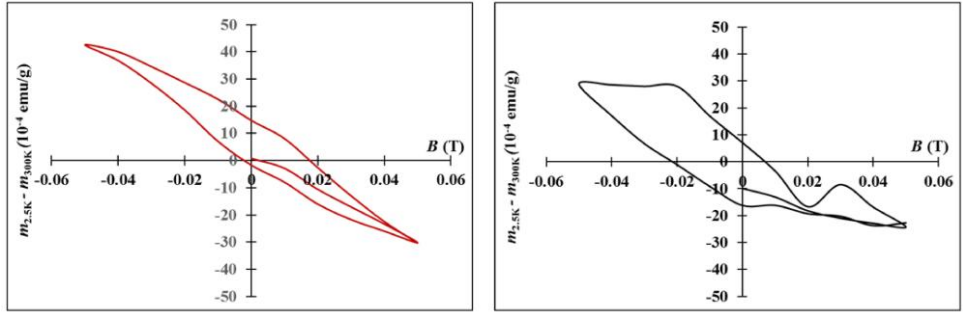

Figure 9: The difference between the magnetic moment at $T$ = 2.5 K and the one at $T$ = 300 K for the H-C foil (left) and the C foil (right).

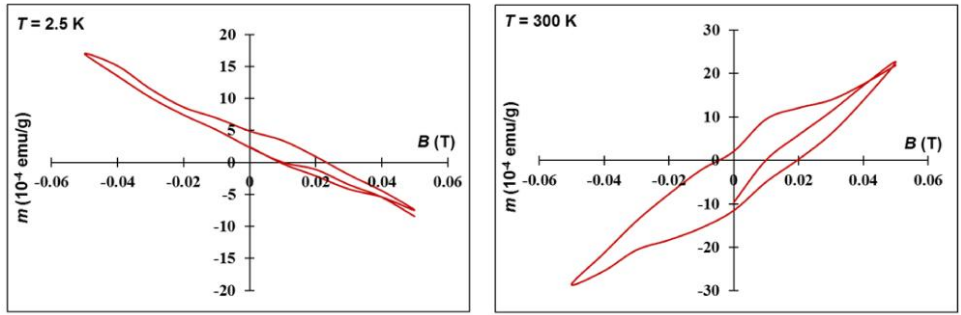

Figure 10: Magnetic moment of the H-C foil less the magnetic moment for the C foil at $T$ = 2.5 K (left) and at $T$ = 300 K (right), respectively.

We should also mention that, in order to observe the unaltered effects of octane on the C foil, the $m(B)$ loops were taken without prior flux cleaning (i.e., degaussing). Nevertheless, a degaussing of the H-C sample was done and here shown in Fig. 11.

The hydrogenation of the C foil done by intercalation with octane showed an interesting time behavior. As the C foil was kept in octane for a longer period of time, the FM fraction increased as a result of longer time exposure to etching (Fig. 12). It is clear that in time, the H-C foils become less DM and more FM.

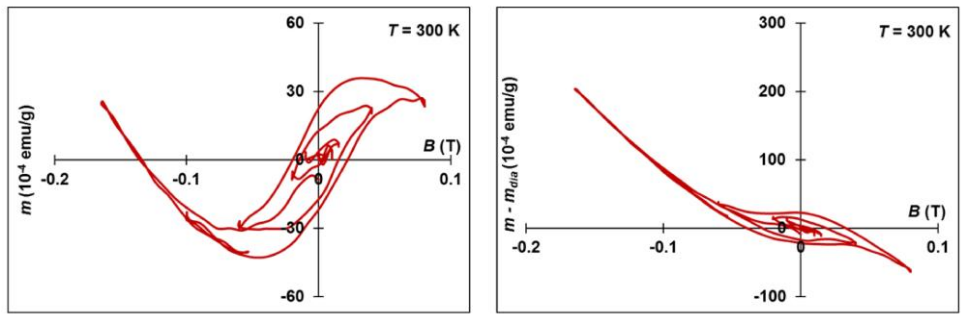

Figure 11: Degaussing of the H-C foil. The magnetic field $H$ was ramped down to zero in an oscillatory manner at a rate -100 Oe/s, from 0 to -1000 Oe, to 800 Oe, to -600 Oe, to 400 Oe, to -200 Oe, to 0. The trend is as expected because $m$ is linear in $H$ (or $B$). The data is shown both with (left) and without (right) DM background, respectively.

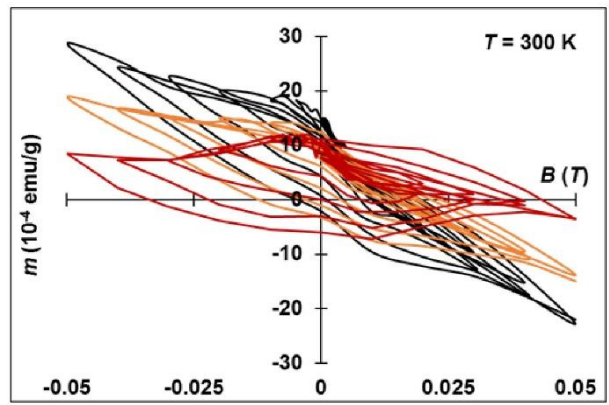

Figure 12: The effect of octane-intercalation period on the H-C foil. Legend: seven months (red) vs. thirteen months (orange) vs. no intercalation (black).

## 4. Conclusions

What is the nature of HTSC in these H-C materials? Following K. Onnes' discovery of SC in mercury below $T_c \simeq 4.19$ K [43], experimentalists and theorists as well have been strongly motivated to find and maybe explain the mechanism for new SC materials with higher $T_c$. In 1957, Bardeen, Cooper, and Schrieffer came up with the breakthrough BCS theory, in which the attractive interactions between electrons are explained as due to their interactions with the lattice excitations, i.e., with phonons [44]. Following the discovery of HTSC materials, it became clear that a new mechanism was needed, in which rather *e-e* correlations than *e*-phonon interactions can lead to high $T_c$. In [1], we have discussed the case of HTSC in hydrogenated graphitic fibers on the basis of Little model for HTSC in organic materials [45]. The model considers a molecular arrangement consisting of two parts:

a) a long chain called the "spine" in which the electrons fill various states and may or may not form a conducting system; b) a series of arms or side chains attached to the spine. Under appropriate choice for the molecules which constitute the side chains, the virtual oscillation of charge in these side chains results in *e-e* correlations, some of which are SC. Interestingly, even if the spine by itself is initially an insulator, the addition of side chains can increase the *e-e* interactions to the point where it becomes energetically favorable to enter the SC state by mixing in states of the conduction band. The spine thus transforms from the insulating or semiconducting state directly to the SC metallic state upon the addition of the side chains. In the case of hydrogenated graphitic fibers or the graphite foil here, the treatment with octane results in the free protonation of octane at graphite's interfaces [46]. The resulted superacidic protons (H+) move freely without activation energy on the graphite surface giving rise to HTSC. Thus, when Little model is applied to the H-C foil, one can imagine that the spines are in the C planes with the protons as the arms or side chains. The free protons are shared by all C atoms in the plane to which the arms are connected to, thus the protons mediate *e-e* correlations that need no input energy, i.e., they make the H-C foil SC. A similar phenomenon was found in [47]. Afterwards, ionic liquid gating-induced protonation was used to obtain hydrogen-doped titanium diselenide showing coexisting superconductivity and charge-density wave [48]. Thus, due to the off-diagonal long-range order that distinguished the SC from normal or insulating states, it is possible that Wigner' search for a quantum system to reproduce itself would lead to a SC as the only system where the probability for such an event would be nonzero. Wigner crystallization, which can be established in a pressurized twisted bilayer graphene, can lead to SC correlations [49]. Usually, the formation of a Wigner crystal occurs when the system is under a magnetic field [50]. As we have found, the magnetic field plays an important role in establishing

SC correlations in hydrogenated graphites and other C-based materials [1, 13, 14, 15].

The results presented here on hydrogenated graphite foil are in agreement with our previous findings, thus confirming HTSC in hydrogenated graphitic materials with critical temperature for the main phase at $T_c \sim 50 - 60$ K. While upping the strength of the magnetic field, the temperature-dependent magnetization reveals several important transitions: from a Neel PM-AFM transition, to a FMSC state, to an orbital PM glass HTSC. The magnetization loops show the kink feature that is characteristic to granular SC, in addition, FM extending up to room temperature. Significantly, the topologically protected flat energy bands in H-C materials [1] promote surface SC of $T_c$ significantly higher than in the bulk. Thus, it appears that when *e-e* correlations and magnetism walk in, these H-C organic materials might just "forge revolutionary paths in the quest for high-temperature superconductivity" [51].

## ACKNOWLEDGMENTS

The experimental part of this work was supported by The Air Force Office of Scientific Research (AFOSR) for the LRIR #14RQ08COR & LRIR #18RQCOR100 and the Air Force Research Laboratory within the Aerospace Systems Directorate (AFRL/RQ). Nadina Gheorghiu acknowledges George Y. Panasyuk for his continuous support and inspiration that made possible this publication.

### References

[1] N. Gheorghiu, C.R. Ebbing, and T.J. Haugan, *Superconductivity in Hydrogenated Graphites*, **arXiv:2005.05876** (2020).

[2] A.C. Basaran, J.E. Villegas, J.S. Jiang, A. Hoffmann, and I.K. Schuller, *Mesoscopic magnetism and superconductivity*, MRS Bulletin **40**, 92 (2015).

[3] M.S. Dresselhaus & G. Dresselhaus, *Intercalation compounds of graphite*, Advances in Physics **51(1)**, 1-186 (2002).

[4] Y. Kopelevich and P. Esquinazi, *Ferromagnetism and Superconductivity in Carbon-based Systems*, J. Low Temp. Phys. **146(5/6)**, 629 (2007).

[5]  P. Esquinazi, A. Setzer, R. Hohne, and C. Semmelhack, *Ferromagnetism in oriented graphite samples*, Phys. Rev. B **66**, 024429 (2002).

[6]  S. Moehlecke, Y. Kopelevich, and M.B. Maple, *Local Superconductivity and Ferromagnetism Interplay in Graphite−Sulfur Composites*, Brazilian Journal of Physics **33(4)**, 762 (2003).

[7]  Y. Kopelevich, R.R. da Silva, J.H.S. Torres, S. Moehlecke, M.B. Maple, *High-temperature local superconductivity in graphite and graphite-sulfur composites*, Physica C **408410**, 77-78 (2004).

[8]  I. Felner, *Peculiar Magnetic Features and Superconductivity in Sulfur Doped Amorphous Carbon*, Magnetochemistry **2**, 34 (2016).

[9]  A. Novac, V.V. Nguyen, E. Fischer, V.G. Lascu, and S.Q. Le, *A stable resistive transition in Y-Ba-Cu-O above 200 K*, Mat. Sci. Eng. **B34**, 147 (1995).

[10] N. Gheorghiu, C.R. Ebbing, and T.J. Haugan, *Electric-field induced strange metal states and possible high-temperature superconductivity in hydrogenated graphitic fibers*, **arXiv:2005.07885** (2020).

[11] V.A. Kulbachinskii, *Coexistence of 2D weak localization and superconductivity in carbon fibers*, Synthetic Metals **41-43**, 2697 (1991).

[12] O.V. Yazyev, *Emergence of magnetism in graphene materials and nanostructures*, Rep. Prog. Phys. **73** 056501 (2010).

[13] N. Gheorghiu N, C.R. Ebbing, B.T. Pierce, and T.J. Haugan, *Quantum effects in graphitic materials: Colossal magnetoresistance, Andreev reflections, ferromagnetism, and granular superconductivity*, IOP Conf. Ser.: Mater. Sci. Eng. **756** 012022 (2020).

[14] N. Gheorghiu, C.R. Ebbing, J.P. Murphy, B.T. Pierce, T.J. Haugan, *Superconducting-like and magnetic transitions in oxygen-implanted diamond-like and amorphous carbon films, and in highly oriented pyrolytic graphite*, **arxiv: 2108.07417**, accepted for publication in Conf. Ser.: Mater. Sci. Eng. (2021-2022).

[15] N. Gheorghiu, C.R. Ebbing, and T.J. Haugan, *Boron Content and the Superconducting Critical Temperature of Carbon-Based Materials*, **arXiv:2012.06624** (2020).

[16] Y. Kubozono et al., *Superconductivity in aromatic hydrocarbons*, Physica C **514**, 199 (2015).

[17] A. Tagliacozzo, G. Campagnano, D. Giuliano, P. Lucignano, and B. Jouault, *Thermal transport driven by charge imbalance in graphene in a magnetic field close to the charge neutrality point at low temperature: Nonlocal resistance*, Phys. Rev. B **99**, 155417 (2019).

[18] I. Grigorenko and R.Ya. Kezerashvili, *Superfluidity of electron-hole pairs between two critical temperatures*, Int. J. Mod. Phys. **29(27)**, 1550188 (2015).

[19] V. Bayot, L. Piraux, J.-P. Michenaud, and J.-P. Issi, *Weak localization in pregraphitic carbon fibers*, Phys. Rev. B **40(6)**, 3514 (1989).

[20] A. Bhaumik *et al.*, *Room-Temperature Ferromagnetism and Extraordinary Hall Effect in Nanostructured Q-Carbon: Implications for Potential Spintronic Devices*, ACS Appl. Nano Mater. **1**, 807 (2018).

[21] N.B. Kopnin, T.T. Heikkilä and G.E. Volovik, *High-temperature surface superconductivity in topological flat-band system*, Phys. Rev B **83**, 220503(R) (2011).

[22] D. Giambastiani, F. Colangelo, A. Tredicucci, S. Roddaro, and A. Pitanti, *Electron localization in periodically strained Graphene*, J. Appl. Phys. **131**, 085103 (2022).

[23] A. Druzhinin, I. Ostrovskii, Yu. Khoverko, N. Liakh-Kaguy & V. Troshina, *Magneto-transport properties of $Bi_2Se_3$ whiskers: superconductivity and weak localization*, Molec. Cryst. Liq. Cryst. **701(1)**, 82 (2020).

[24] G. Baskaran, *Resonating-valence-bond contribution to superconductivity in $MgB_2$*, Phys. Rev. B **65**, 212505 (2002).

[25] D.A. Khokhlov and R.S. Akzyanov, *Pauli paramagnetism of triplet Cooper pairs in a nematic superconductor*, Phys. Rev. B **104**, 214514 (2021).

[26] Y. Yeshurun, I. Felner, and H. Sompolinsky, *Magnetic properties of a high-$T_c$ superconductor $YBa_2Cu_3O_7$: Reentrylike features, paramagnetism, and glassy behavior*, Phys. Rev. B **36(1)**, 840 (1987).

[27] E.R. Podolyak, *On Magnetic Flux Trapping by Surface Superconductivity*, JEPT **126(3)**, 389 (2018).

[28] A.K. Geim, S.V. Dubonos, J.G.S. Lok, M. Henini, & J.C. Maan, *Paramagnetic Meissner effect in small superconductors*, Nature **396**, 144 (1998).

[29] L.A.B. Olde Olthof, X. Montiel, J.W.A. Robinson, and A.I. Buzdin, *Superconducting vortices generated via spin-orbit coupling at superconductor/ferromagnet interfaces*, Phys. Rev. B **100**, 220505(R) (2019).

[30] W. Braunisch *et al.*, *Paramagnetic Meissner effect in high-temperature superconductors*, Phys. Rev. B **48(6)**, 48 (1993).

[31] D. Domínguez, E.A. Jagla and C. A. Balseiro, *Wohlleben Effect in a model granular High TC Superconductor*, Physica C **235-240**, 3283 (1994).

[32] W.A. Ortiz, P.N. Lisboa-Filho, W.A.C. Passos, F.M. Araujo-Moreira, *Field-induced networks of weak-links: an experimental demonstration that the paramagnetic Meissner effect is inherent to granularity*, Physica C **361**, 267 (2001).

[33] S.-I. Suzuki and Y. Asano, *Paramagnetic instability of small topological superconductors*, Phys. Rev. B **89**, 184508 (2014).

[34] S.-I. Suzuki, T. Hirai, M. Eschrig, and Y. Tanaka, *Anomalous inverse proximity effect in unconventional superconductor junctions*, Phys. Rev. Res. **3**, 043148 (2021).

[35] E. Lake and T. Senthil, *Reentrant superconductivity through a quantum Lifshitz transition in twisted trilayer graphene*, Phys. Rev. B **104**, 174505 (2021).

[36] G.E. Volovik, *Graphite, graphene and the flat band superconductivity*, **arXiv:1803.08799v1** (2018).

[37] S. Layek *et al.*, *Possible high temperature superconducting transitions in disordered graphite obtained from room temperature deintercalated $KC_8$*, **arXiv:2205.09358v1** (2022).

[38] F.V. Kusmartsev, *Orbital Glass in HTSC: a New State of Condensed Matter*, J Supercond Nov Magn **5(5)**, 463 (1992).

[39] N. Garcia and P. Esquinazi, *Mean-Field Superconductivity Approach in Two Dimensions*, J Supercond Nov Magn **22**, 439 (2009).

[40] S. Kumar, R.P. Singh, A. Thamizhavel, C.V. Tomy, *Evidence of surface superconductivity and multi-quanta vortex state in a weakly-pinned single crystal of $Ca_3Ir_4Sn_{13}$*, Physica C **509**, 42 (2015).

[41] W.A.C. Passos *et al.*, *Granularity in superconductors: intrinsic properties and processing-dependent effects*, Physica C **354**, 189 (2001).

[42] D. Saint-James and P.G. deGennes, *Onset of superconductivity in decreasing fields*, Phys. Lett. **7(5)**, 30 (1963).

[43] H.K. Onnes, Commun. Phys. Lab. Univ. Leiden. **120b**, **122c**, **122d** (1911).

[44] J. Bardeen, L. N. Cooper, and J. R. Schrieffer, *Theory of Superconductivity*, Phys. Rev. **108**, 1175 (1957).

[45] W.A. Little, *Possibility of Synthesizing an Organic Superconductor*, Phys. Rev. A **134(6A)**, 1416 (1964).

[46] Y. Kawashima & M. Iwamoto, *Protolytic decomposition of n-octane on graphite at near room temperature*, Sci. Rep. **6**, 28493 (2016).

[47] Y. Kawashima, *Superconductivity above 500 K in conductors made by bringing n-alkane into contact with graphite*, **arXiv:1612.05294** (2016).

[48] E. Piatti *et al.*, *Coexisting superconductivity and charge-density wave in hydrogen-doped titanium diselenide via ionic liquid gating-induced protonation*, **arXiv: 2205.12951v1**, (2022).

[49] B. Padhi and P.W. Phillips, *Pressure-induced metal-insulator transition in twisted bilayer graphene*, Phys. Rev. B **99**, 205141 (2019).

[50] Yu P. Monarkha and V.E. Syvokon, *A two-dimensional Wigner crystal (review article)*, Low Temp. Phys. **38**, 1067 (2012).

[51] D. Jerome, *Organic Superconductors: When Correlations and Magnetism Walk in*, J Supercond Nov Magn **25**, 633 (2012).